\documentclass{aa}  

\usepackage{graphicx}
\usepackage{subcaption}
\usepackage{amsmath}
\usepackage{multirow}

\usepackage{txfonts}
\newcommand{\asec}{^{\prime\prime}}

\begin{document}

   \title{Evidence for a massive dust-trapping vortex connected to spirals}

   \subtitle{A multi-wavelength analysis of the HD~135344B protoplanetary disk}

   \author{P. Cazzoletti
          \inst{1}
          \and
          E. F. van Dishoeck
          \inst{1,2}
          \and
          P. Pinilla
          \inst{3}
          \and 
          M. Tazzari
          \inst{4}
          \and
          S. Facchini
          \inst{1}
           \and
          N. van der Marel
          \inst{5}
          \and
          M. Benisty
          \inst{6,7}
          \and
          A. Garufi
          \inst{8}
          \and
          L. M. P\'erez
          \inst{9}
          }

   \institute{Max-Planck-Institut f{\"u}r Extraterrestrische Physik, Gießenbachstraße, 85741 Garching bei München, Germany\\
              \email{pcazzoletti@mpe.mpg.de}
               \and
             Leiden Observatory, Leiden University, Niels Bohrweg 2, 2333 CA Leiden, The Netherlands
             \and Department of Astronomy/Steward Observatory, The University of Arizona, 933 North Cherry Avenue, Tucson, AZ 85721, USA
             \and Institute of Astronomy, University of Cambridge, Madingley Road, Cambridge CB3 0HA, UK
             \and Herzberg Astronomy \& Astrophysics Programs, National Research Council of Canada, 5071 West Saanich Road, Victoria BC V9E 2E7, Canada
		\and Unidad Mixta Internacional Franco-Chilena de Astronom\'{i}a (CNRS, UMI 3386), Departamento de Astronom\'{i}a, Universidad de
Chile, Camino El Observatorio 1515, Las Condes, Santiago, Chile
\and Univ. Grenoble Alpes, CNRS, IPAG, 38000 Grenoble, France. 
		\and INAF, Osservatorio Astrofisico di Arcetri, Largo Enrico Fermi 5, I-50125 Firenze, Italy
		\and Departamento de Astronom\'ia, Universidad de Chile, Camino El Observatorio 1515, Las Condes, Santiago, Chile
             }

   \date{Received September 15, 1996; accepted March 16, 1997}

 
  \abstract
   {Spiral arms, rings and large scale asymmetries are structures observed in high resolution observations of protoplanetary disks, and it appears that some of the disks showing spiral arms in scattered light also show asymmetries in millimeter-sized dust. HD~135344B is one of these disks. Planets are invoked as the origin of these structures, but no planet has been observed so far and upper limits are becoming more stringent with time.}
   {We want to investigate the nature of the asymmetric structure in the HD~135344B disk in order to understand the origin of the spirals and  of the asymmetry seen in this disk. Ultimately, we aim at understanding whether or not one or more planets are needed to explain such structures.}
   {We present new ALMA sub-$0.1\asec$ resolution observations at optically thin wavelengths ($\lambda = 2.8$ mm and $1.9$ mm)  of the HD~135344B disk.  The high spatial resolution allows us to unambiguously characterize the mm-dust morphology of the disk. The low optical depth of continuum emission probes the bulk of the dust content of the vortex. Moreover, we combine the new observations with archival data at shorter wavelengths to perform a multi-wavelength analysis and to obtain information about the dust distribution and properties inside the observed asymmetry.}
   {We resolve the asymmetric disk into a symmetric ring + asymmetric crescent, and observe that: (1) the spectral index strongly decreases at the center of the vortex, consistent with the presence of large grains; (2) for the first time, an azimuthal shift of the peak of the vortex with wavelength is observed; (3) the azimuthal width of the vortex decreases at longer wavelengths, as expected for dust traps. These features allow to confirm the nature of the asymmetry as a vortex. Finally, under the assumption of optically thin emission, a lower limit to the total mass of the vortex is $0.3 M_{\rm Jupiter}$. Considering the uncertainties involved in this estimate, it is possible that the actual mass of the vortex is higher and possibly within the required values ($\sim 4\,\rm M_{\rm Jupiter}$) to launch spiral arms similar to those observed in scattered light. If this is the case, no outer planet is needed to explain the morphology.}
   {}

   \keywords{protoplanetary disks - planet-disk interaction - planets and satellites: formation - stars: individual (HD~135344B) - instabilities               }

   \maketitle
%

\section{Introduction}

In recent years, as higher angular resolution observations of protoplanetary disks are routinely performed, it is becoming apparent that almost every single disk hosts some structures. Rings \citep[e.g.][]{2016ApJ...820L..40A,2016PhRvL.117y1101I, 2018A&A...610A..24F}, cavities and large-scale asymmetries \citep[e.g.][]{2013Natur.493..191C,2013Sci...340.1199V} and spirals \citep[e.g.][]{2015A&A...578L...6B,2017A&A...597A..42B, 2016Sci...353.1519P,2017AJ....153..264F,2016A&A...595A.113S,2017ApJ...849..143S, 2018ApJ...860..124D} are being observed at mm-wavelength continuum emission with ALMA and in scattered light in the near-infrared (NIR) with VLT/SPHERE \citep{2008SPIE.7014E..18B} and Gemini/GPI \citep{2008SPIE.7015E..18M}.

Rings and azimuthal asymmetries are often interpreted as dust traps, i.e. local pressure maxima where dust grains marginally coupled to the gas remain trapped \citep{1977MNRAS.180...57W}. Such traps have critical implications in planet formation, since they allow dust particles to grow to large sizes and eventually to form planetesimals by preventing them from rapidly drifting into the central star \citep{1997Icar..128..213K,2007Natur.448.1022J,2012A&A...538A.114P,2012A&A...545A..81P}.

\begin{figure*}[t]
  \includegraphics[width=\textwidth]{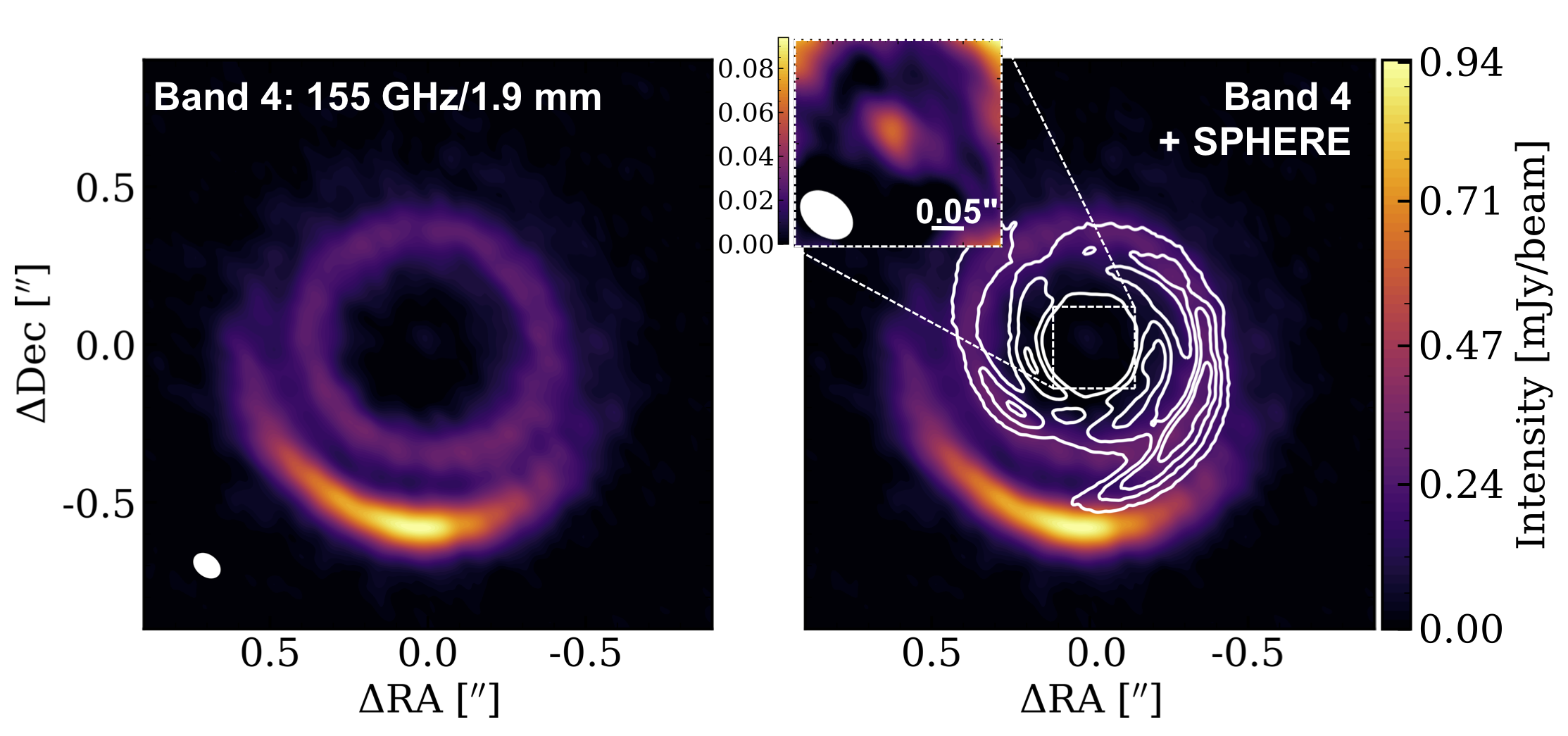}
\caption{\textit{Left}: ALMA Band 4 observations of HD~135344B at 1.9 mm. The panel in the upper right corner shows a zoom into the inner regions, with a colormap enhancing the emission from the central point-source. \textit{Right}: same image with the contours of the spiral arms observed with SPHERE \citep{2017ApJ...849..143S} overlaid in white. Note that the spiral starts near the vortex position. The white ellipse in the bottom left corner of the left panel shows the beam size. }\label{fig:data_b4}
\end{figure*}

\begin{figure}
  \includegraphics[width=0.5\textwidth]{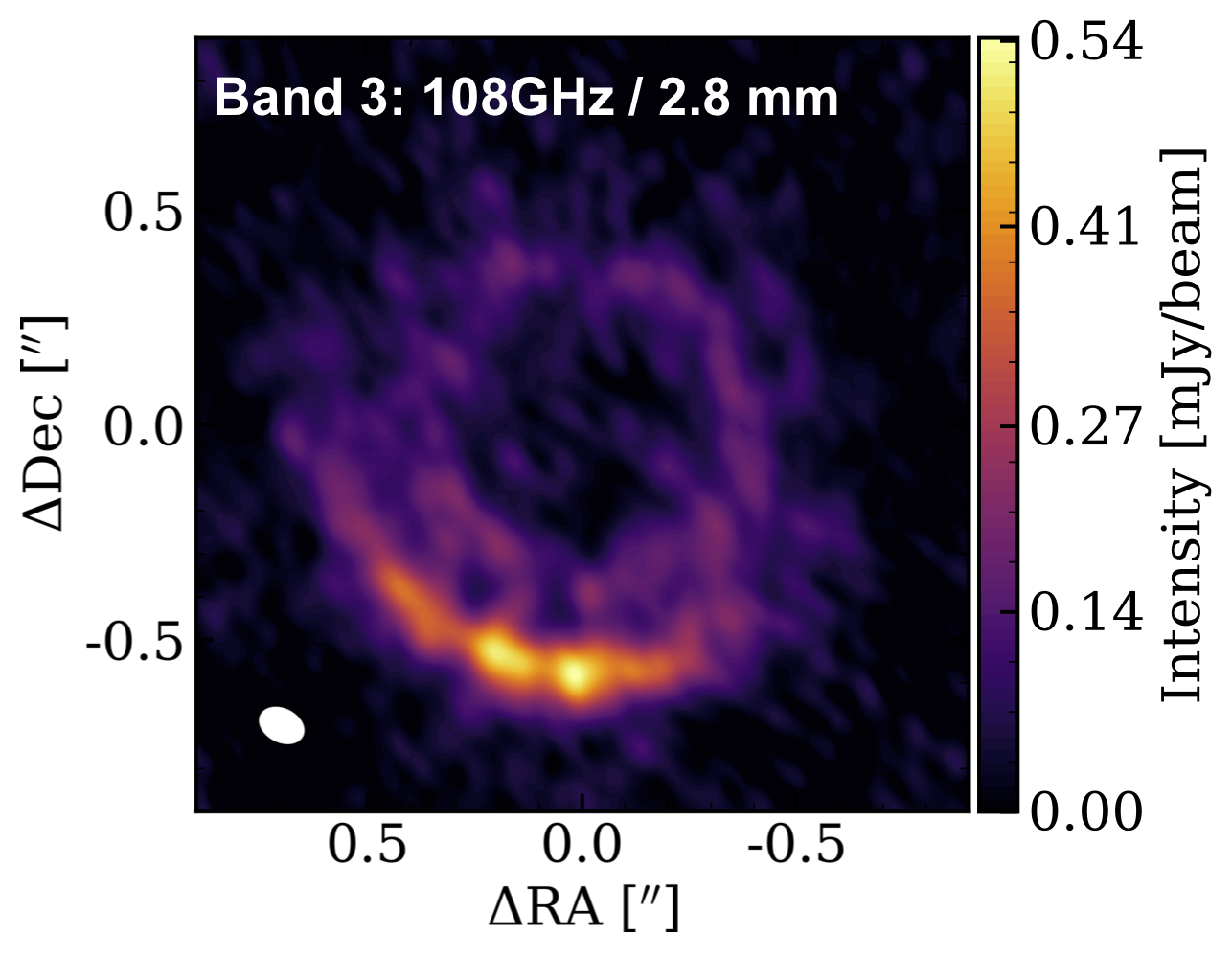}
\caption{Band 3 observations of HD~135344B at 2.8 mm. The white ellipse in the bottom left corner shows the beam size.}\label{fig:data_b3}
\end{figure}

Planets are invoked as a potential origin for pressure bumps, and can also explain the observed spiral arms as these are the result of an exchange of angular momentum between the planet and the material in the disk \citep[e.g.][]{2015ApJ...809L...5D,2015ApJ...815L..21F,2016ApJ...819..134B,2017ApJ...835...38D}. Symmetric, ring-shaped dust traps are expected to form when a planet is massive enough \citep[e.g.][]{dipierro,2016MNRAS.459.2790R}. The edges of these gaps can then become unstable against Rossby wave instability  \citep[RWI,][]{1999ApJ...513..805L,2000ApJ...533.1023L,2001ApJ...551..874L} giving rise to vortices,  i.e., high-pressure regions that can trap dust azimuthally and form crescent-shaped dust asymmetries such as those observed at mm-wavelength \citep[e.g.][]{2009A&A...493.1125L,2012MNRAS.426.3211L} . The number of planets and their location are, however, degenerate, and the same system can often be explained by more than one scenario.  Moreover, the lack of directly imaged planets and the increasingly strong upper limits on their mass \citep[e.g.][$ M_{\rm planet}<4\,\rm M_{\rm Jupiter}$  at $ r\geq0.6\asec$ in HD~135344B]{2017A&A...601A.134M} are challenging our understanding of the origin of such structures\footnote{These upper limits are calculated assuming a hot start scenario; in the case of a cold start the upper limits could be substantially higher.}.

The wealth of information provided by observations at different wavelengths can now help us to break some of the degeneracies in the interpretation of the observed morphologies. In particular, to understand the role of planets in sculpting the disk morphology and vice versa, to understand the role of  structures in the planet formation process, it is now timely  to look for a common explanation for the different morphologies observed for grains of different sizes. 

In this context, HD~135344B, also known as SAO~206462, is a particularly interesting object. Located at a distance of $135\pm1.4$ pc \citep{2016A&A...595A...1G, 2018arXiv180409366L, 2018arXiv180409376L,2018arXiv180409378G}, it is a transition disk with an inner mm-dust cavity of $\sim 40$ au \citep{2007ApJ...664L.107B,2009ApJ...704..496B,2011ApJ...732...42A}. In addition, the presence of a deep $^{13}$CO and C$^{18}$O gas cavity whose size is smaller than that of the dust \citep{vdm2016_isot} suggests that the inner cavity observed in the dust could have been carved by a large inner planet at a radius $\lesssim30$ au from the central star \citep[as also suggested by e.g. ][]{2011AJ....142..151L}.

\begin{figure}
  \includegraphics[width=0.5\textwidth]{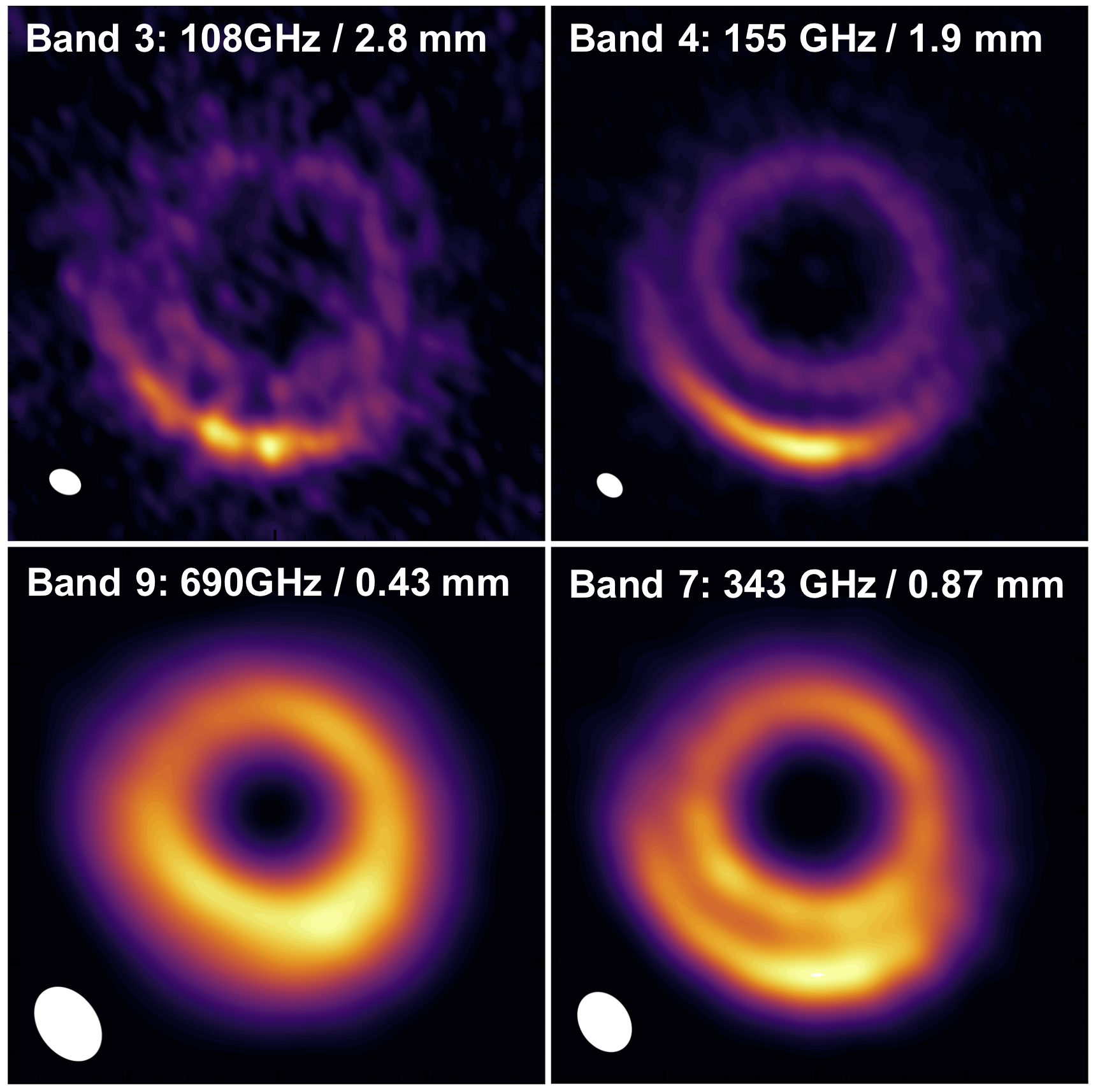}
\caption{Observations of HD~135344B at different wavelengths. The white circle has a radius of $0.6\asec$. The white ellipses show the beam size of each observation.}\label{fig:data_all_wavelengths}
\end{figure}

At the same time, this system shows two bright, symmetric spiral arms in scattered light \citep{2012ApJ...748L..22M,2013A&A...560A.105G,2016A&A...595A.113S,2017ApJ...849..143S} extending out to $\sim75$ au, that can be explained either by two planets at 55 and 126 au \citep{2016A&A...595A.113S} or by a single, massive planet outside the dust cavity at $\sim100$ au \citep{2015ApJ...815L..21F,2016ApJ...819..134B,2017ApJ...835...38D}. However, no planet has yet been observed. Using $0.2\asec$ resolution data, \citet{vdm2016}  noted that the mm emission could be due to an inner symmetric ring and an outer vortex, rather than a single asymmetric ring. They proposed that the vortex may be triggering one of the observed spiral arms, while the massive inner planet, also carving the dust cavity, may be launching the second arm. No planet other than the inner one would therefore be required in this explanation. 

Structures similar to those in HD~135344B, i.e. rings and asymmetries in the mm continuum and spirals arms in scattered light, have subsequently been observed in V1247~Ori \citep{2017ApJ...848L..11K} and MWC~758 \citep{2017ApJ...840...60B,2018ApJ...860..124D}, strengthening the idea of a connection between the asymmetries observed in the (sub-)mm continuum emission and spiral arms in scattered light. 

\begin{figure}
\centering
\includegraphics[width=0.5\textwidth]{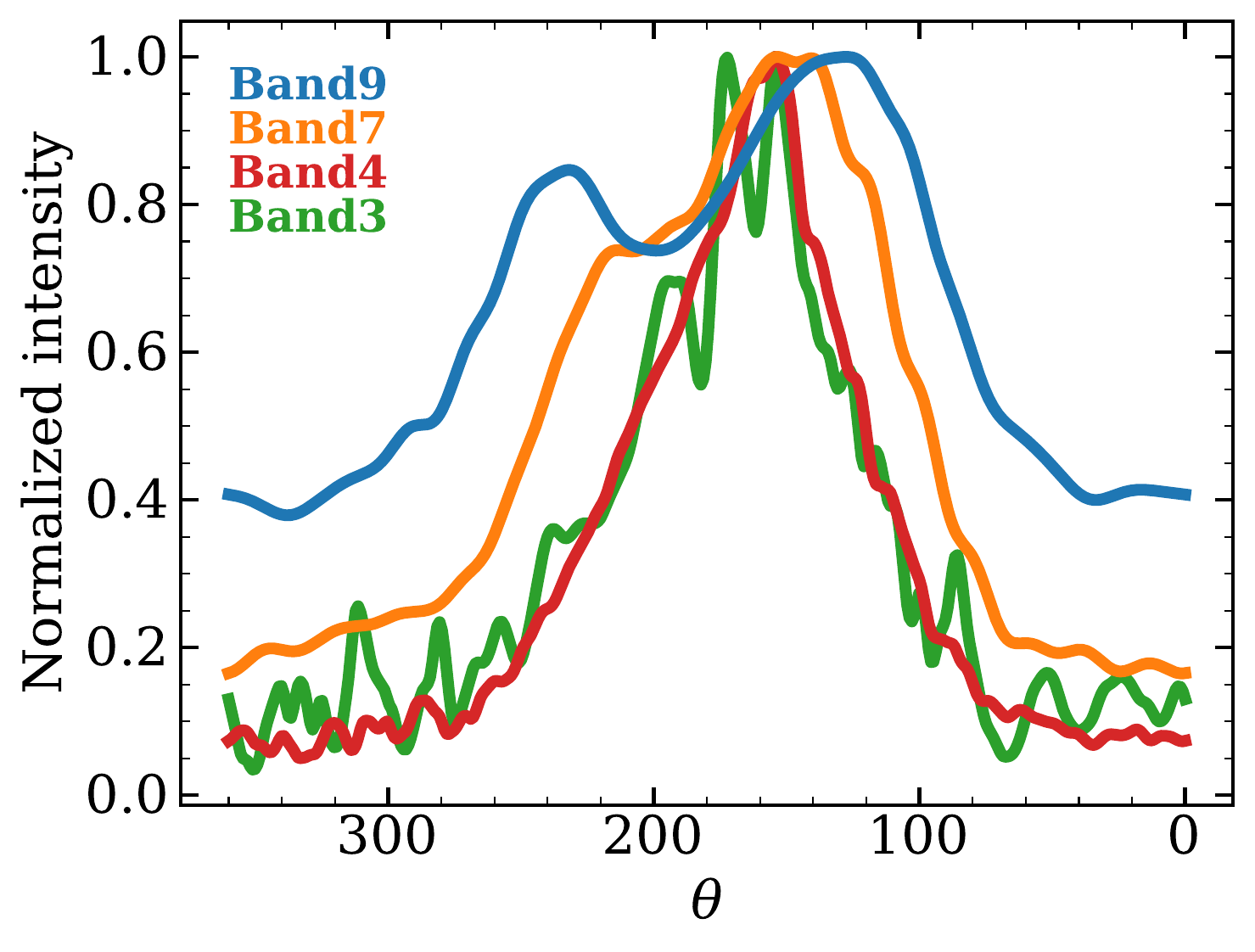}
\caption{Normalized intensities at different wavelengths as a function of the azimuth, measured along the asymmetry (i.e. at radius $r=0.6\asec$) after deprojecting the data. Note that the intensity peak shifts monotonically  to smaller values of $\theta$ as the wavelength increases.}\label{fig:azi_cuts}
\end{figure}

The origin of the structures observed in HD~135344B is however still ambiguous, both because of the limited spatial resolution of the available data and because azimuthal asymmetries can in principle be due to mechanisms different than vortices \citep[e.g. disk eccentricity, as in][]{2013A&A...553L...3A, 2017MNRAS.464.1449R} which cannot launch spiral density waves. 

Spiral arms can also be explained by other mechanisms not involving an external perturber. Symmetrical spiral arms resembling the scattered light morphology of the HD~135344B disk can in principle arise from gravitational instability \citep[GI, e.g.][]{2015MNRAS.451..974D,2015ApJ...812L..32D}, although for this specific system the total disk mass does not seem to be high enough for GI to arise. Shadows have also been proposed to trigger spiral arms \citep{2016ApJ...823L...8M}, and have been observed in the disk around HD~135344B. Such shadows, however, change on a much faster time-scale than the spirals.

In this work, we present new high-resolution Cycle 4 and Cycle 5 ALMA observations of the HD~135344B disk at 1.9 mm and 2.8 mm, which allows us to test the different scenarios. We reach a much higher resolution of $0.08\asec$. The dust emission is expected to be optically thin at these long wavelengths, and the bulk of the dust mass inside the observed structures can be determined. These new observations, combined with archival observations at 0.87 mm and 0.43 mm for the same object, allow us to carry on a detailed multi-wavelength analysis.

In Sec. \ref{sec:obs}, the observations and the data reduction process are discussed. In Sec. \ref{sec:results}, we present the observed morphology and the approach used to fit the data. In Sec. \ref{sec:analysis}, we use the data to obtain information on the dust and gas content of the vortex and to test theoretical predictions on vortices. Finally, in  Sec. \ref{sec:discussion}, we use our new data and results to discuss the origin of the morphology observed in HD~135344B and its role in planet formation.

\section{Observations and data reduction}\label{sec:obs}
The new ALMA data of the disk around HD~135344B presented in this paper are part of three different projects. The Band 4 data have been observed in C40-8 configuration during ALMA Cycle 4 in two execution blocks on September 14th, 2017 and September 28th, 2017 as part of program 2016.1.00340.S (PI: Cazzoletti) for a total of $\sim45$ min with 41 and 42 antennas. Additional Band 4 observations were carried out during Cycle 5 in C43-5 configuration, on January 17th, 2018 as part of the DDT program 2017.A.00025.S (PI: Cazzoletti) in order to recover the short uv-spacings, for a total of $\sim22$ min on source and using 44 antennas.  The Band 3 data were observed in Cycle 5 in the C43-8 configuration on November 11th, 2017 (2017.1.00884.S, PI: Pinilla) for $~\sim 24$ min using 44 antennas, and in the C43-5 configuration on January 17th, 2018 (as part of the DDT proposal 2017.A.00025.S) for $\sim11$ min on source and 46 antennas.

The spectral setups of the Band 4 and 3 observations are as follows. In order to maximize continuum sensitivity, the Band 4 observations have 4 spectral windows each, with  a bandwidth of 1875 MHz (channel width $\sim 2$ MHz, corresponding to $\sim 3.6\,\rm km\,s^{-1}$), with rest frequencies at 161.987722, 159.997750, 149.997890 and 147.997918 GHz. The spectral setup of the Band 3 observations also have 4 spectral windows in total. Two of them are centered on the $^{13}$CO(1-0) and C$^{18}$O(1-0) transitions with rest frequencies at 110.201354 and 109.782176 GHz, and a bandwidth of 937.5 MHz (resolution of 488 kHz) and 234.38 MHz (resolution of 122 kHz) respectively. The other two windows are dedicated to the continuum, and are both centered at 108 GHz. In one of them the correlator is set to Time Division Mode (128 channels, 31.25 MHz resolution and 1875 MHz total bandwidth), while in the other one it is set to Frequency Division Mode (3840 channels, 488 kHz resolution, 937.5 MHz total bandwidth). In all executions,  J1427-4206 was used as bandpass, pointing and absolute flux calibrator, and J1457-3539 was used as phase calibrator.

The archival observations are from ALMA Cycle 0 program
2011.0.00724.S (P.I. P{\'e}rez) and Cycle 1 program
2012.1.00158.S (P.I. van Dishoeck), taken in Band 9 (690 GHz)
and Band 7 (343 GHz), respectively. The details of the calibration are
described in \citet{perez2014} and \citet{vdm2016_isot}.

Data were then processed and calibrated through the ALMA  calibration pipeline in the Common Astronomy Software Applications \citep[CASA 5.1.1,][]{2007ASPC..376..127M}. The high signal-to-noise allows self-calibration of the Band 4 data (both amplitude and phase) after standard phase referencing. The visibilities were then Fourier inverted and deconvolved using the \texttt{TCLEAN} task in \texttt{CASA}, and imaged using Briggs weighting with a Robust parameter of 0.5. The synthetized beam size achieved are $0.09\asec\times0.063\asec$ (Band 4 data) and  $0.1\asec\times0.073\asec$ (Band 3 data). Note that the analysis carried out in this work was performed in the uv-plane only, and is therefore not affected by the cleaning process. The achieved sensitivity in Band 3 and Band 4 is $25.3\,\rm \mu Jy\,beam^{-1}$ and $10.9\,\rm \mu Jy\,beam^{-1}$, respectively. The corresponding peak signal-to-noise is 21 in the Band 3 data, and 86 in Band 4.

\section{Results}\label{sec:results}
\subsection{ALMA images}
The Band 3 (108 GHz, $\lambda= 2.8\,\rm mm$) and Band 4 (155 GHz,  $\lambda= 1.9\,\rm mm$) observations of the HD~135344B disk, imaged as described above, are presented in Fig. \ref{fig:data_b4} and \ref{fig:data_b3}.  The dusty component of the HD~135344B disk consists of a symmetric inner ring and an outer azimuthally asymmetric, crescent-shaped structure, potentially due to a vortex. The ring and the crescent are separated by a gap. The Band 4 data also show a "bridge" at the west end of the crescent, connecting it to the inner ring. Interestingly, the bridge emission is located exactly where the spiral arm seen in scattered light crosses the gap between the ring and the crescent (see also Fig. \ref{fig:data_b4}), although it could also just be due to the synthesized beam elongation.

\begin{figure}
  \includegraphics[width=0.5\textwidth]{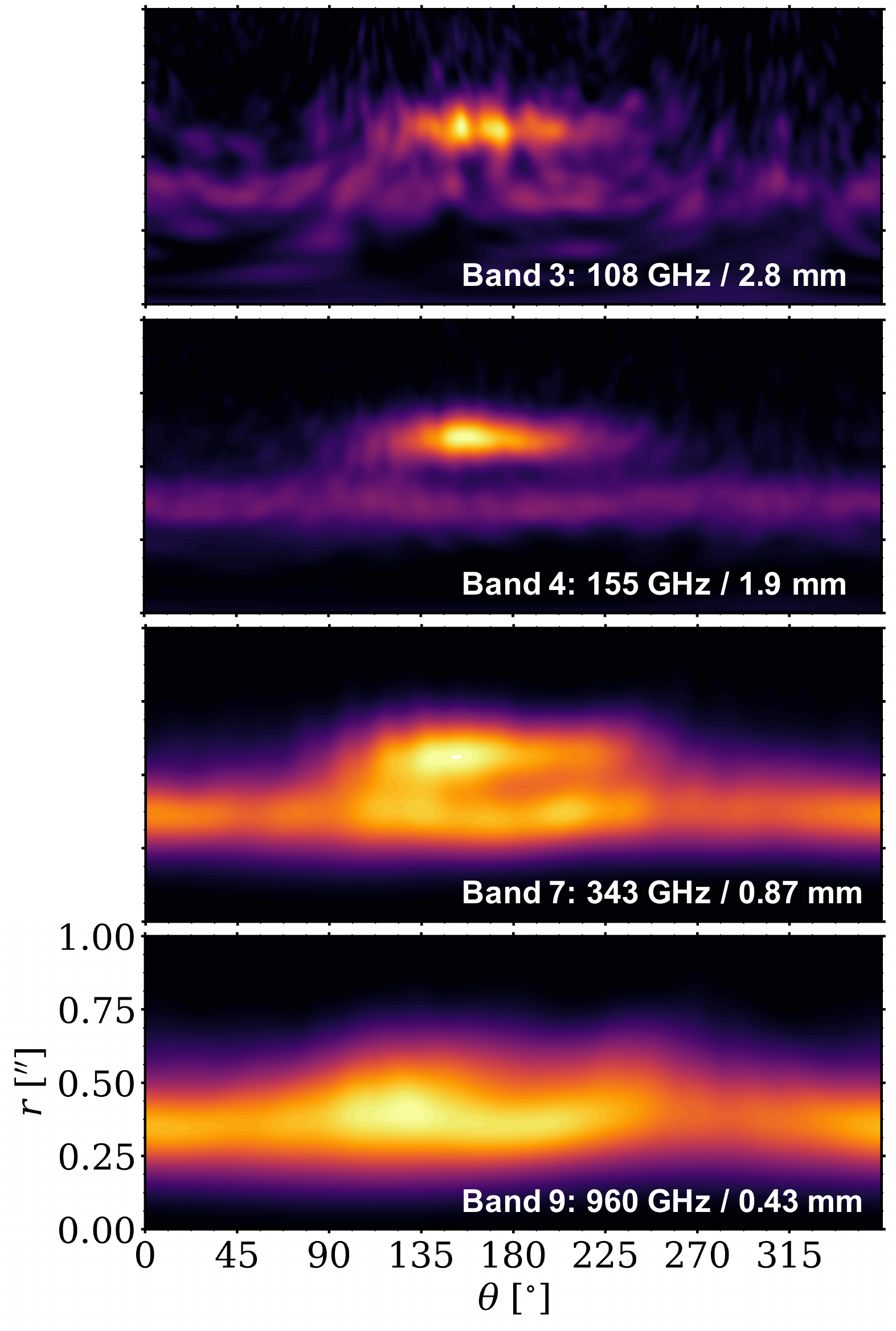}
\caption{Observations of HD~135344B at different wavelengths in polar coordinates. The data have been deprojected using $PA=62^\circ$ and the inclinations in Table \ref{tab:bestfit}. Note the displacement of the peak of the vortex emission between the different wavelengths; also note that the ring has eccentricity very close to zero.}\label{fig:data_polar}
\end{figure}

Although the images have different spatial resolution, a visual comparison between the Band 3 and Band 4 observations at $0.08\asec$ with the archival Band 7 and Band 9 at $0.2\asec$ shows some clear differences (Figs. \ref{fig:data_all_wavelengths} and \ref{fig:data_polar}). First, the relative contrast between the flux of the asymmetry and that of the ring increases with increasing wavelengths, showing that the two components have different spectral indices and perhaps even different dust populations. Second, the images suggest that the emission peak of the crescent is located at different azimuthal locations in the different data sets, and moves toward the east at longer wavelengths, as also highlighted by the azimuthal cuts in Fig. \ref{fig:azi_cuts}.

The high signal-to-noise Band 4 observations also allow for a $\sim5\,\sigma$ detection of a central unresolved source, either due to a small inner circumstellar disk surrounding the star or to free–free/synchrotron emission from ionized gas in the proximity of the star (see right panel in Fig. \ref{fig:data_b4}), which appears to be at the center of the dust ring. The presence of a small inner disk is consistent with the large near infrared excess \citep{2017A&A...603A..21G} due to the presence of hot dust in the innermost regions, and with the shadows observed in scattered light by \citet{2017ApJ...849..143S}. No significant central emission is visible in the Band 3 observations because of the lower signal-to-noise, nor at shorter wavelengths due to the beam dilution (see end of Sec. \ref{sec:mass}).

\subsection{Model fit in the uv-plane}\label{sec:fit}
To quantitatively constrain the large scale structures in the HD~135344B disk we fit the surface brightness in the uv-plane with a simple analytical model similar to that adopted by \citet{perez2014}, \citet{2015A&A...584A..16P} and \citet{vdm2016}. The inner ring is fitted with an azimuthally symmetric Gaussian in the radial direction, centered at a radius $r_{\rm R}$ and with $\sigma=\sigma_{\rm R}$:
\begin{equation}
F_1(r,\theta) =F_{\rm R} e^{-(r-r_{\rm R})^2/2\sigma_{\rm R}^2},
\end{equation}
where $F_{\rm R}$ is the peak surface brightness. The outer asymmetric crescent is modelled as a double Gaussian in both the radial and azimuthal direction, based on the vortex prescription by \citet{2013ApJ...775...17L}, radially centered at $r_{\rm V}$, with a radial width $\sigma_{\rm V,\, r}$, and azimuthally peaking at $\theta_{\rm V}$. As for the azimuthal width, both preliminary tests and visual inspection of our Band 4 data show that the surface brightness of the system is better represented by an azimuthally asymmetric Gaussian rather than a symmetric one. We therefore allow our model to have two different widths $\sigma_{\rm V, \,\theta1}$ and $\sigma_{ \rm V,\,\theta2}$ east and west of the center of the asymmetry, respectively. Our adopted model is therefore:
\[ 
F_2(r,\theta) = 
\begin{cases} 
      F_{\rm V} e^{-(r-r_{\rm V})^2/2\sigma_{\rm V, \,r}^2}e^{-(\theta-\theta_{\rm V})^2/2\sigma_{\rm V, \,\theta1}^2} & \theta \leq \theta_{\rm V} \\
      F_{\rm V} e^{-(r-r_{\rm V})^2/2\sigma_{\rm V, \,r}^2}e^{-(\theta-\theta_{\rm V})^2/2\sigma_{\rm V, \,\theta2}^2} & \theta > \theta_{\rm V} 
   \end{cases}
\]

$F_{\rm V}$  is the peak surface brightness of the asymmetry at location $(r_{\rm V},\theta_{\rm V})$. The final model $F(r,\theta)=F_1(r,\theta)+F_2(r,\theta)$ uses 9 parameters.

The parameters are then found by Fourier-transforming the analytical model, by sampling it at the same uv-locations as the observations using the code \texttt{GALARIO} \citep{tazzari2018}, and by then fitting them to the observed visibilities with the Markov-chain-Monte Carlo (MCMC) code \texttt{emcee} \citep{2013PASP..125..306F}. The center of the ring $(x_{\rm c}, y_{\rm c})$ and the inclination $i$ are also left as 3 additional free parameters, while the position angle $PA$, defined east-of-north, is fixed to $62^{\circ}$, the value found by \citet{vdm2016_isot} from the CO isotopologues moment 1 maps. The Band 3 and Band 4 data, as well as the archival Band 7 and Band 9 data sets, are fitted independently with the model described above. In the MCMC, 120 walkers were used and the parameters were extracted using 24000 likelihoods.

\begin{figure}
\centering
\includegraphics[width=\linewidth]{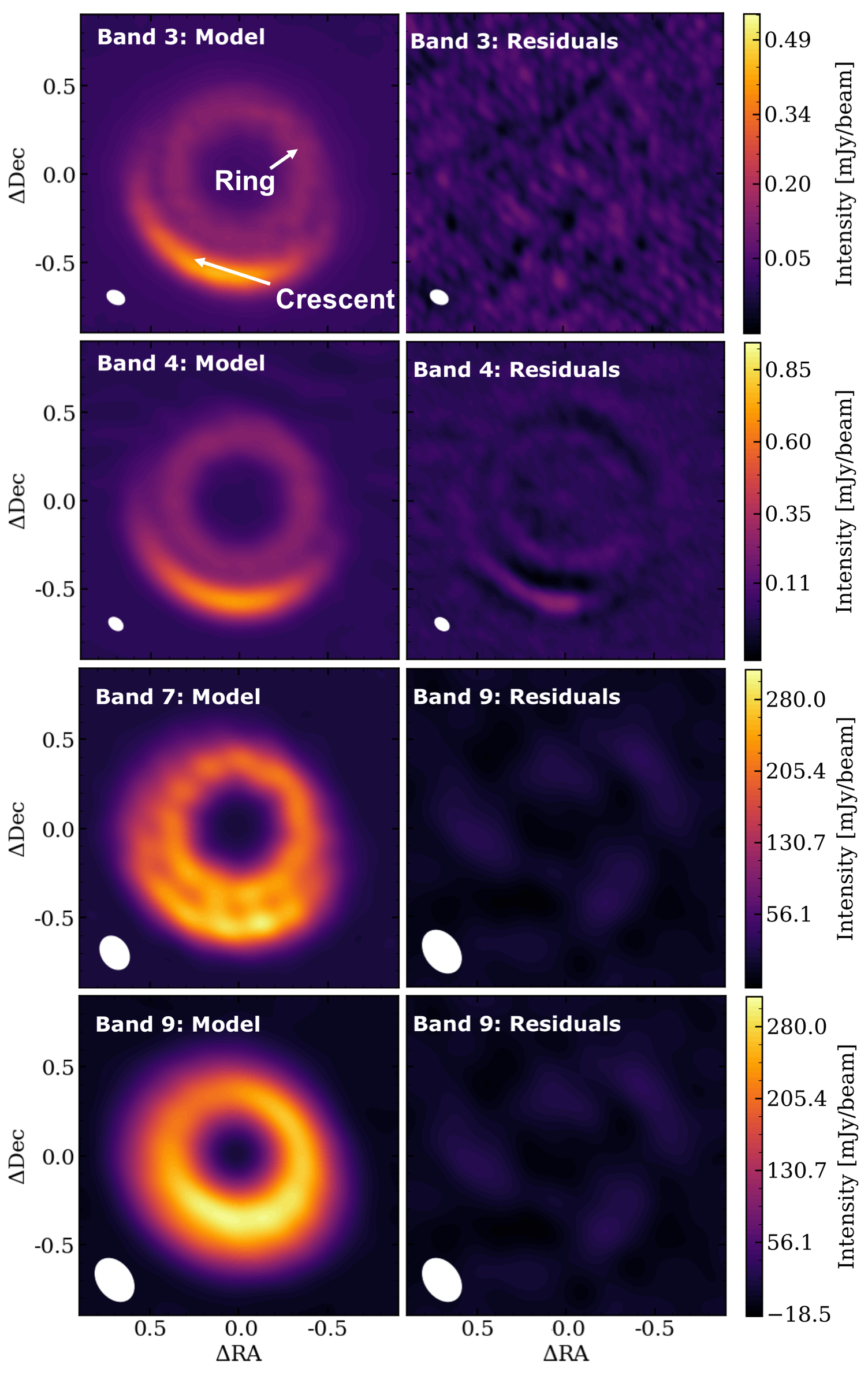}
\caption{Models derived from the MCMC using the parameters in Tab. \ref{tab:bestfit}, and relative residuals for the four data sets available for HD~135344B. The color scale is the same for the models and the residuals. The white ellipses show the synthesized beam sizes.}\label{fig:models}
\end{figure}

The derived parameters at each wavelength are presented in Table \ref{tab:bestfit}. The model shows good agreement with the data at all wavelengths, with the inner ring located at $0.38\asec$ ($\sim50$ au) radius and the crescent at $0.6\asec$ ($\sim80$ au) from the central star.

Fig. \ref{fig:models} shows the synthesised images of the obtained models and relative residuals, derived with the same imaging parameters (mask, weighting and number of iterations) as the data. The data are generally very well fit, and residuals of at most 25\% at the peak are found only in the Band 4 data set. These are due to the finest resolved structures that cannot be well represented by a simple double Gaussian as that used to model the crescent.

A comparison between the real and imaginary part of the visibilites of the data and the models are shown in Appendix \ref{sec:appendix_visibilities}.

\begin{table*}[th]
\renewcommand*{\arraystretch}{1.4}
\caption{Parameters estimated from the marginalized likelihood distribution. The value is estimated as the median. The upper and lower uncertainty represent the extent to the 16th and 84th percentile, respectively.}
\label{tab:bestfit}
\centering
\begin{tabular}{lcccccc}
\hline\hline
\multicolumn{3}{}{}  & Band 3	 &  Band 4 & Band 7 & Band 9 \\
\multicolumn{3}{}{}  & 108 GHz	 &  155 GHz & 343 GHz & 690 GHz \\
\hline
\multirow{ 3}{*}{Ring} & $\log F_{\rm R}$ & [Jy/sr] & $8.71 ^{+0.03}_{-0.03}  $ & $9.229 ^{+0.004}_{-0.004} $ & $10.621 ^{+0.002}_{-0.002} $ & $11.464 ^{+0.005}_{-0.004} $ \\
& $r_{\rm R} $& [$\asec$] & $0.381 ^{+0.006}_{-0.007} $ & $0.3752 ^{+0.0009}_{-0.0009} $ & $0.3686 ^{+0.0003}_{-0.0003} $ & $0.354 ^{+0.001}_{-0.001} $ \\
& $\sigma_{\rm R}$ & [$\asec$] & $0.097 ^{+0.009}_{-0.008} $ & $0.0858 ^{+0.0009}_{-0.0011} $ & $0.0623 ^{+0.0004}_{-0.0004} $ & $0.070 ^{+0.001}_{-0.001} $ \\
\hline
\multirow{ 6}{*}{Crescent} & $\log F_{\rm V}$ & [Jy/sr] & $9.36 ^{+0.02}_{-0.02}  $ & $9.758 ^{+0.003}_{-0.003}  $ & $10.747 ^{+0.002}_{-0.002} $ & $11.306 ^{+0.006}_{-0.007} $ \\
& $r_{\rm V} $& [$\asec$] & $0.603 ^{+0.009}_{-0.009}$ & $0.587 ^{+0.002}_{-0.001}$ & $0.5783 ^{+0.0003}_{-0.0003} $ & $0.544 ^{+0.002}_{-0.002} $ \\
& $\sigma_{\rm r, V}$ & [$\asec$] & $0.058 ^{+0.003}_{-0.003}$ & $0.0545 ^{+0.0005}_{-0.0005} $ & $0.0626 ^{+0.0004}_{-0.0004} $ & $0.085 ^{+0.001}_{-0.001} $ \\
& $\theta_{\rm V} $& [$\asec$] & $162.1 ^{+2.5}_{-2.3} $ & $153.6 ^{+0.4}_{-0.4} $ & $137.4 ^{+0.1}_{-0.1} $ & $113.9 ^{+1.1}_{-1.0} $ \\
& $\sigma_{\theta1, \rm V}$ & [$\asec$] & $34.9 ^{+2.2}_{-1.9} $ & $33.8 ^{+0.4}_{-0.4} $ & $34.6 ^{+0.2}_{-0.2} $ & $41.7 ^{+0.9}_{-0.8} $ \\
& $\sigma_{\theta2, \rm V} $& [$\asec$] & $40.9 ^{+2.8}_{-2.6} $ & $47.8 ^{+0.4}_{-0.4} $ & $77.0 ^{+0.3}_{-0.2} $ & $109.6^{+1.9}_{-1.8} $ \\
\hline
& $i$ & [$^\circ$] & $17.7 ^{+3.4}_{-4.7} $ & $11.9 ^{+0.8}_{-0.7} $ & $9.8 ^{+0.1}_{-0.1} $ & $11.6 ^{+0.8}_{-0.8} $ \\
& $x_{\rm c}$ &  & 15:15:48.417$^{+0.006\asec}_{-0.006\asec}$ & 15:15:48.417$^{+0.001\asec}_{-0.001\asec}$ & 15:15:48.419$^{+0.0003\asec}_{-0.0003\asec}$ & 15:15:48.423$^{+0.0007\asec}_{-0.0007\asec}$ \\
& $y_{\rm c}$ &  & -37.09.16.433$ ^{+0.007\asec}_{-0.008\asec}$& -37:09:16.453$^{+0.001\asec}_{-0.001\asec}$ & -37.09.16.346$^{+0.0003\asec}_{-0.0003\asec}$ & -37.09.16.333$^{+0.0007\asec}_{-0.0007\asec}$ \\
\hline
\end{tabular}
\end{table*}

\section{Analysis}\label{sec:analysis}
\subsection{Dust mass}\label{sec:mass}
Under the assumption of optically thin emission it is possible to convert the observed flux into dust mass. In particular, the flux density measured in the crescent is $8.6\,\rm mJy$ at 108 GHz, and $20.2\,\rm mJy$ at 155 GHz. The dust mass is then calculated as
\begin{equation}
M_{\rm dust}=\frac{F_\nu d^2}{\kappa_\nu B_\nu(T_{\rm dust})},
\label{eq:mass}
\end{equation}
where the distance $d=136\,\rm pc$, the dust temperature $T_{\rm dust}$ at the location of the asymmetry (i.e. $\sim80$ au) is assumed to be 15 K, as modelled by \citet{vdm2016_isot} using radiative transfer to fit the spectral energy distribution, and the dust grain opacity $\kappa_\nu$ is taken as $10\,\rm cm^2 g^{-1}$ at $1000\,\rm GHz$ and scaled to the observed frequencies using an opacity power-law index $\beta=1$ \citep{1990AJ.....99..924B}. The calculated dust masses are $44\,\rm M_{\oplus}$ and $52\,\rm M_{\oplus}$ in Band 4 and Band 3, respectively. 

These numbers should be regarded as lower limits. First, mm-wavelength observations are insensitive to the emission from very large pebbles (cm-size and larger) which, if present, may be hiding part of the mass.  Moreover, the dust densities reached at such high dust masses can make even mm-wavelength emission partially optically thick. An estimate of the optical thickness can be obtained from the brightness temperature $T_{\rm bright}$. The calculated peak brightness temperatures are 7 K in Band 3 and 8 K in Band 4. The relation between brightness temperature and physical temperature $T_{\rm dust}$ is $T_{\rm bright} = T_{\rm dust}(1-\exp{(-\tau_\nu)})$, which when inverted (using again $T_{\rm dust}$ = 15 K) gives an optical thickness $\tau_\nu$ of $\sim 0.5-0.6$ at 108 and 155 GHz. We conclude that the emission can only be considered marginally optically thin, and that the dust mass of the asymmetry is likely higher than $50\,\rm M_{\oplus}$. The crescent remains only marginally optically thin even assuming $T_{\rm dust}$ higher by a factor of 1.5,  with $\tau_\nu\sim 0.4$.

The inferred dust surface density at the peak of the asymmetry, derived from the relation $\tau_\nu=\Sigma_{\rm dust}\kappa_\nu$ for $\tau_\nu=0.6$, and using the same value of $\kappa_\nu$ as above, is of the order of $\sim1\,\rm g\,cm^{-2}$.

The total flux densities in Band 3 and Band 4 are $\sim20.0\,\rm mJy$ and ${\sim42.7\,\rm mJy}$, respectively. The dust masses corresponding to these flux densities, calculated using Eq. \ref{eq:mass} and assuming $T_{\rm dust}=15\,\rm K$, are $\sim90\,\rm M_{\oplus}$ in Band 4 and $\sim120\,\rm M_{\oplus}$ in Band 3. About half of the dust is therefore concentrated inside the asymmetry.

The unresolved emission at the star location, seen in Band 4, has a flux density of $\sim85\,\rm\mu Jy$, i.e., assuming it is due to a small inner disk, its dust mass is $\sim0.1 \,\rm M_\oplus$. The expected flux density in Band 7 can be calculated with $F_{343\,\rm GHz}=F_{155 \,\rm GHz}\times(343/155)^2$ (assuming Rayleigh-Jeans, optically thick emission) which gives $F_{343\,\rm GHz} = 416\,\rm \mu Jy$, which is well below $3\sigma$ at the Band 7 rms level. The inner ring is therefore not observable in the current shorter wavelength observations.

\begin{figure}
\centering
\includegraphics[width=0.5\textwidth]{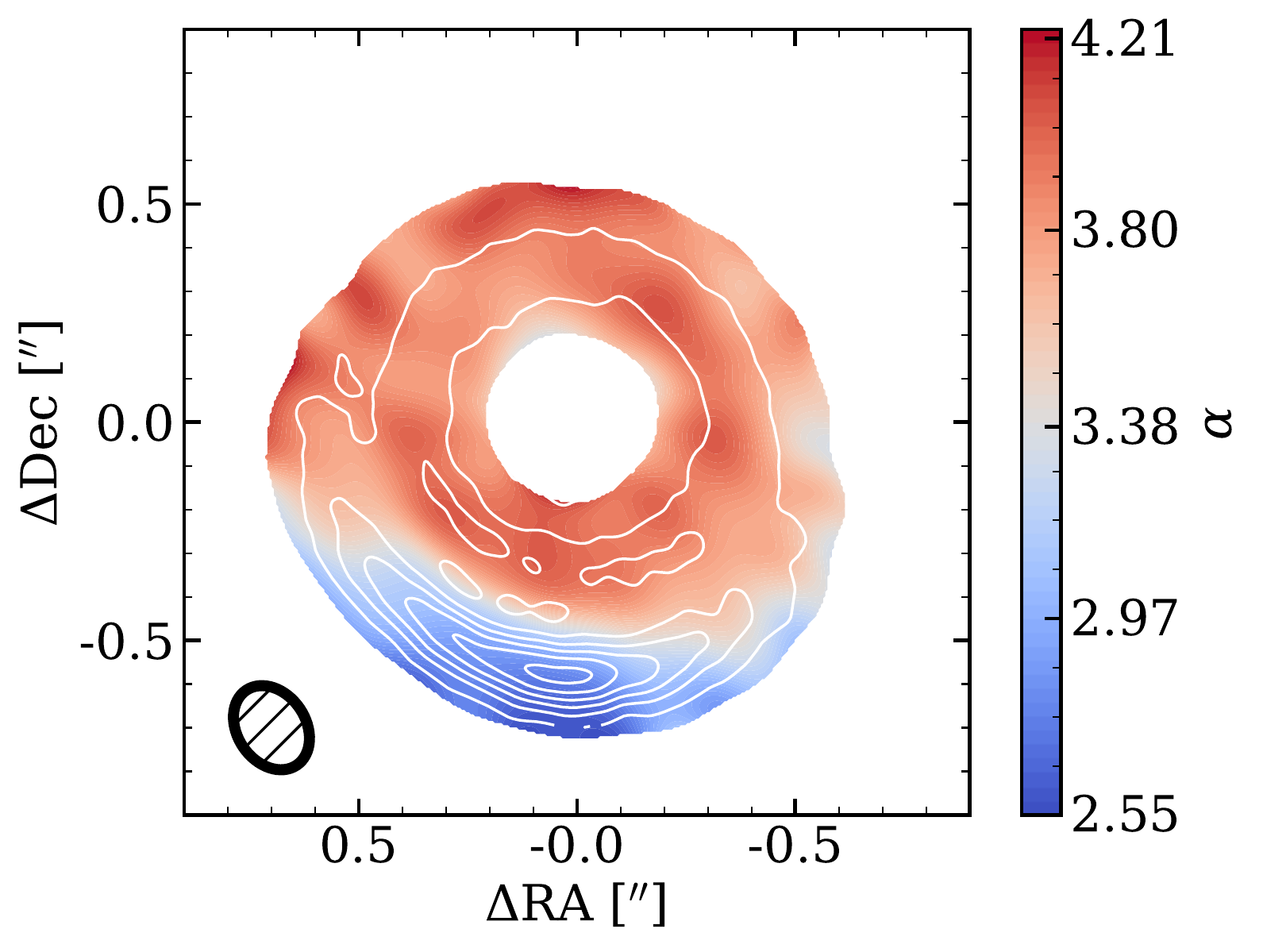}
\caption{Map of the spectral index in HD~135344B, calculated between the Band 4 and the Band 7 data. A clear azimuthal variation in the value of $\alpha$ can be seen, consistent with the presence of large grains trapped inside a vortex. The typical uncertainty is $\pm0.3$. White contours show the Band 4 intensity profile at $0.08\asec$ resolution to highlight the location of the crescent. The ellipse shows the beam size of the Band  4 and Band 7 maps.}\label{fig:alpha}
\end{figure}

\subsection{Spectral index distribution}
In a particle trap, dust rapidly reaches larger sizes and grain growth can be measured through the spectral index of the dust opacity $\beta$. In the Rayleigh-Jeans regime, the spectral index $\alpha$ of the spectral energy distribution $F_\nu\propto\nu^\alpha$ at mm-wavelengths can be related to $\beta$, that can be interpreted in terms of grain size when the dust continuum emission is optically thin \citep[e.g.][]{2006ApJ...636.1114D,2014prpl.conf..339T}. Values of $\alpha$ with $2<\alpha<3$ hint at the presence of large, cm-sized grains and hence dust growth. In the case of an azimuthal trap, clear spatial variations of the inferred grain size along the azimuthal direction are expected, with the smallest values near the center of the vortex. Therefore, spatial variations of $\alpha$ should be detected. In this section we calculate the spatial distribution of $\alpha$ in the HD~135344B disk.

The spectral index is calculated between Band 4 and Band 7: the choice  of these two wavelengths turned out to be the best trade-off between large enough wavelength leverage, high signal-to-noise, and enough spatial resolution to confidently calculate the value of $\alpha$. uv-tapering is applied to the Band 4 data set, in order to obtain the same synthesized beam as that of Band 7. Finally, the spectral index $\alpha$ is calculated as
${\alpha=\ln(F_{\rm 155\, GHz}/F_{\rm 335\, GHz})/\ln({\rm 155\, GHz/ 335\, GHz})}$, after re-centering the images on the best-fit center obtained in section \ref{sec:fit}. The value of $\alpha$ is calculated using only those pixels with a signal-to-noise greater than 20 in both bands. The uncertainty is therefore dominated by the $\pm5\%$ and $\pm10\%$ calibration uncertainties in Band 4 and Band 7 respectively, translating in a $\pm0.15$ uncertainty on the absolute value of $\alpha$. The obtained map is shown in Fig. \ref{fig:alpha}.

\begin{figure*}[]
\centering
\begin{subfigure}{0.43\textwidth}
\centering
\includegraphics[page=1,width=1\linewidth]{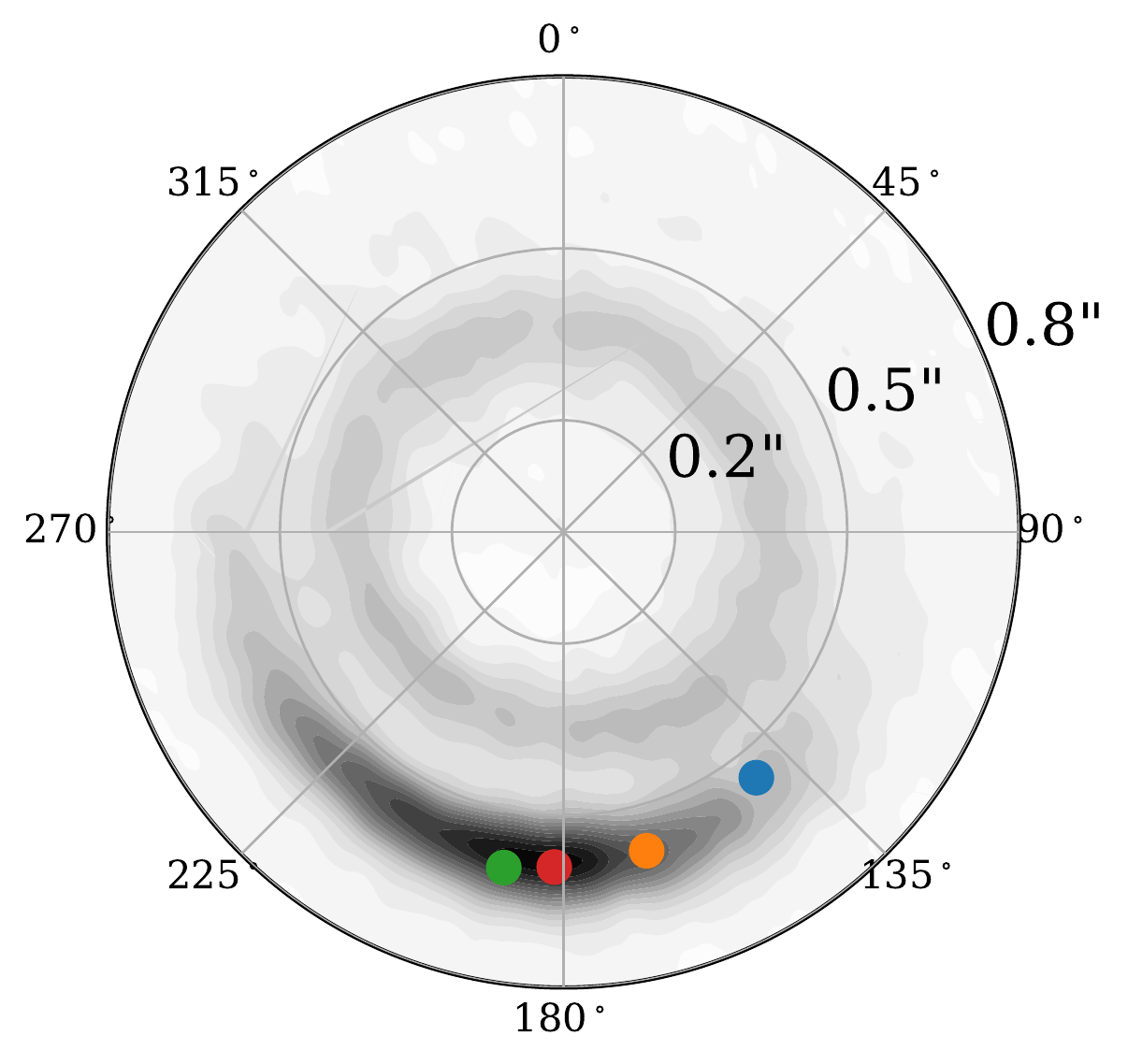}
\phantomcaption\label{fig:peaks_data} 
\end{subfigure}
\begin{subfigure}{0.56\textwidth}
\centering
  \includegraphics[width=1\textwidth]{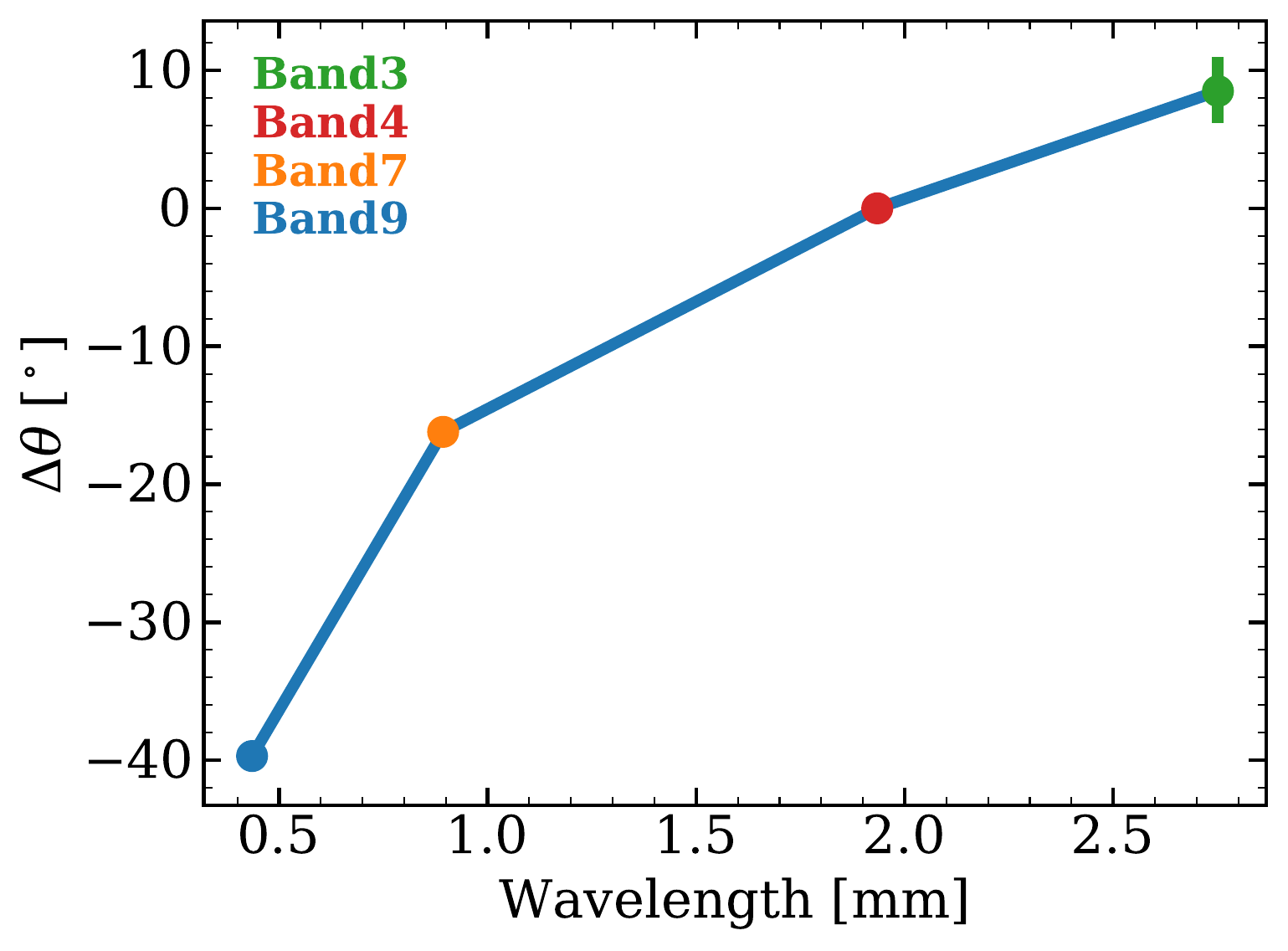}
\phantomcaption\label{fig:peaks_s_wl}
\end{subfigure}
\caption{\textit{Left}: Location of the peak of the emission at different wavelengths as obtained from the fit. The grey contours show the Band 4 data.  \textit{Right}: azimuthal shift of the emission peak at different wavelengths, calculated with respect to the position of the Band 4 emission peak. There is a shift of more than $30^\circ$ between the Band 3 and Band 9 emission. The uncertainty on the peak location in Band 4, Band 7 and Band 9 is smaller than the size of the dot.}\label{fig:peak_shift}
\end{figure*}

As expected, there is a clear azimuthal variation in the value of $\alpha$,  with $\alpha$ decreasing to values below 3 inside the crescent. This is consistent with the presence of large dust grains, and therefore with dust growth, but can also be due to high optical thickness, or to a combination of both effects. Interestingly, the symmetric ring appears to be both optically thin and devoid of large grains, with $\beta=1.8\pm0.3$ consistent with interstellar medium (ISM) grains. The measured variation across the map is not affected by the absolute calibration uncertainty.

\subsection{Peak shift}\label{sec:peakshift}

The fit of the brightness profile allows to quantitatively compare the morphology of the emission inside the crescent at different wavelengths. In particular, as shown in Fig. \ref{fig:peak_shift}, we observe that at longer wavelengths the fitted peak of the crescent emission is azimuthally shifted towards the east, as already qualitatively seen in Fig. \ref{fig:azi_cuts}. Keplerian rotation can immediately be ruled-out as the cause of such a shift, as the gas-kinematics combined with the scattered light (which allows to identify the close and far sides of the disk) show that the rotation is in the opposite direction. Assuming that the emission at longer wavelengths is dominated by larger grains, and in particular that grain size $\sim3\times\lambda$ \citep{2006ApJ...636.1114D}, we observe a shift of $\sim50^\circ$ between the 1.5 mm grains (traced in Band 9) and 9 mm grains (traced in Band 3).

A different distribution between small and large dust grains has been predicted by \citet{2015ApJ...798L..25M,2016MNRAS.458.3927B} in vortices, and is an effect due to the vortex gas self-gravity. However, the direction of the shift is just the opposite in these models, which always predict a shift ahead of the vortex rather than behind as in the case of the HD~135344B disk. \citet{2016MNRAS.458.3927B}, in particular, show that, whereas smaller grains are trapped close to the center of the gas vortex, larger grains can get trapped ahead of it. The amount of shift between the large and small grains depends on the coupling between the gas and the dust, parametrized with the Stokes number ${\rm St}$ defined as
\begin{equation}
\mbox{{\rm St}}=\frac{\pi}{2}\frac{s\rho_{\rm pc}}{\Sigma_{\rm gas}}\approx0.015\times\Bigg(\frac{s}{1\,\rm mm}\Bigg)\Bigg(\frac{\rho_{\rm pc}}{1\,\rm g\, cm^{-3}}\Bigg)\Bigg(\frac{10\,\rm g\,cm^{-2}}{\Sigma_{\rm gas}}\Bigg),
\label{eq:st}
\end{equation}
where $s$ is the radius of the dust grains, $\rho_{\rm pc}$ is their internal density, $\Sigma_{\rm gas}$ is the gas surface density. In practice, particles with ${\rm St}>>1$ are decoupled from the gas, while those with ${\rm St}<<1$ are well coupled. Trapping usually occurs when ${\rm St}\lesssim1$.

An azimuthal shift as large as that observed, between grains which are only a factor of a few different in size, means that the observed mm-sized grains are neither strongly coupled nor very decoupled from the gas, and that their Stokes number must be close to unity. Assuming a typical value of $\rho_{\rm pc}=1\rm \, g\, cm^{-3}$, we can therefore invert Eq. \ref{eq:st} and obtain an estimate of the gas surface density at the vortex location of $\Sigma_{\rm gas}\approx1\,\rm g\,cm^{-2}$. This gas surface density is consistent with that obtained by \citet{vdm2016_isot} by modelling the emission from $^{13}$CO and C$^{18}$O.  The absolute value of the displacement observed in our data is consistent also with the simulations by \citet{2016MNRAS.458.3927B} and indeed corresponds to grains with ${\rm St}\sim1$. However, the fact that their models cannot reproduce the direction of the shift observed in HD~135344B indicates that more simulations exploring a larger parameter space or studying different dynamical effects are still needed.

Even though the derivation of the local gas surface density has large uncertainties in both the Stokes number method and for the CO isotopologue one, it can be combined with the estimate of 
$\Sigma_{\rm dust}$ based on the optical thickness of our Band 4 data to get a qualitative estimate of the gas-to-dust ratio. We find that the gas-to-dust ratio inside the asymmetry is close to unity, i.e. significantly lower than 100, the value generally assumed for interstellar clouds. Given the large uncertainties involved in our calculations, our estimate is in line with the gas-to-dust ratio of 10 commonly found in simulations of dust traps \citep[e.g.][]{2017ApJ...835..118M, 2017ApJ...850..115S}. It should be noted that when the gas-to-dust ratio approaches 1, dust feedback becomes relevant and can even cause dynamical instabilities potentially destroying the vortex \citep{2014ApJ...795L..39F}. The simulations by  \citet{2016MNRAS.458.3927B} do not include dust feedback.

\subsection{Azimuthal dust trapping}
Simulations of dust dynamics and evolution in vortices also predict a different azimuthal concentration for differently sized grains \citep{birnstiel2013}. In particular, one of the key predictions in dust traps is that smaller grains with lower Stokes number, traced at shorter wavelengths,  tend to be less concentrated than larger grains observed at longer wavelengths. This behaviour is expected only when dust-trapping is responsible for the observed crescent-shaped structure, while it is not present when the asymmetry is due to other mechanisms such as disk eccentricity. So far, this effect has been observed for the vortex in the Oph IRS~48 disk \citep{2015ApJ...810L...7V}, for the HD~142527 disk \citep{2015ApJ...812..126C}, and the MWC~758 disk \citep{2015ApJ...813...76M,2018arXiv180503023C}.

\begin{figure}
\centering
\includegraphics[width=0.5\textwidth]{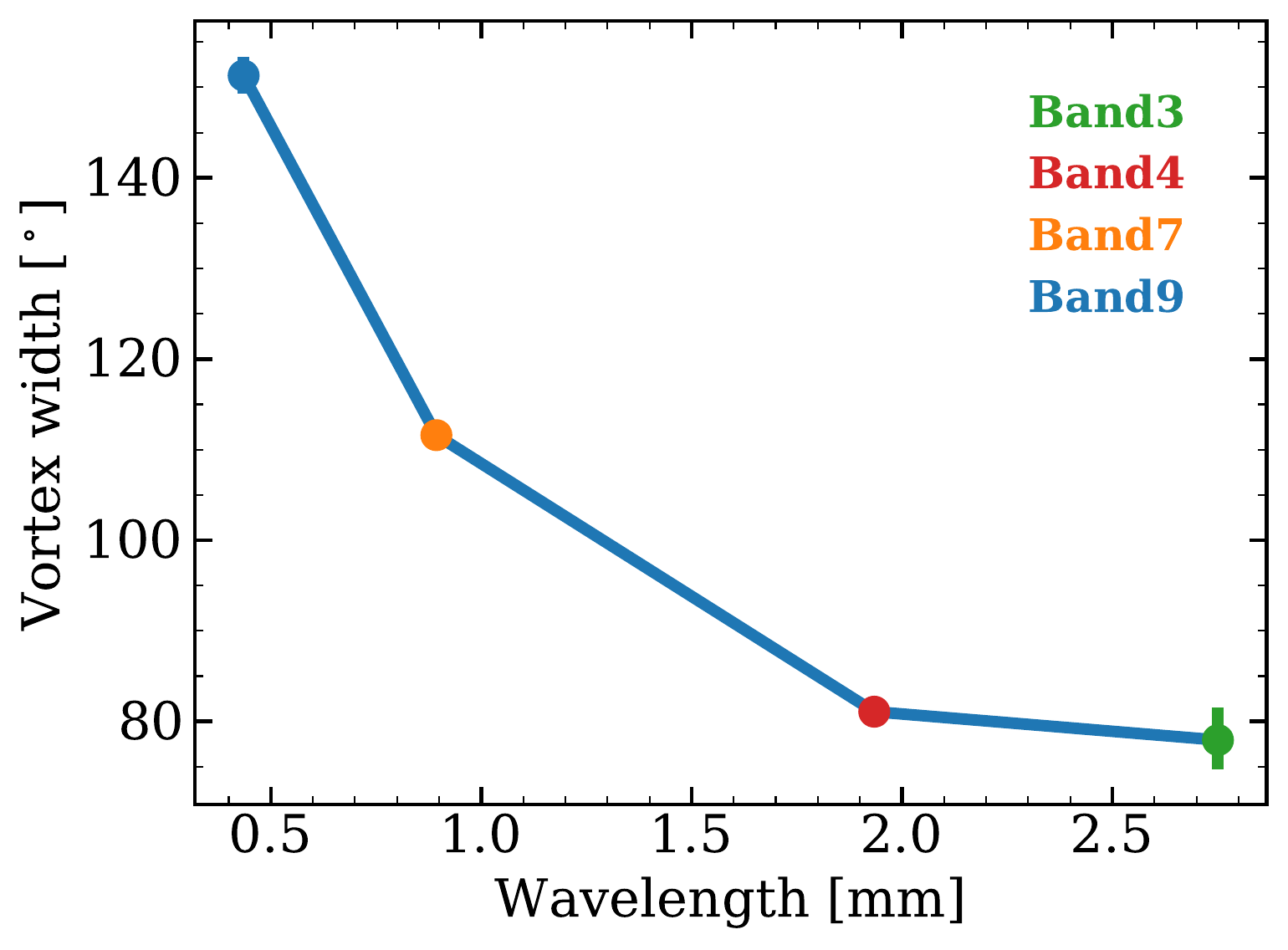}
\caption{Azimuthal width of the crescent as a function of wavelength. Larger particles are more azimuthally concentrated than small particles, that are more coupled to the gas. The error bars in Band 4 and Band 7 are smaller than the size of the dots.}\label{fig:width}
\end{figure}

We define the total width of the asymmetry as ${\sigma_{\theta1\,\rm,V}+\sigma_{\theta2\,\rm ,V}}$. Using the values obtained for the models at all wavelengths (Tab. \ref{tab:bestfit}), the azimuthal width is found to monotonically decrease with wavelength, as shown in Fig. \ref{fig:width}, consistent with the theoretical expectation for dust-trapping.

\section{Discussion}\label{sec:discussion}
In Section \ref{sec:analysis} different theoretical predictions for dust traps were tested, leading to the identification of the crescent-shaped structure in the HD~135344B disk as a vortex, trapping mm-sized dust grains. Dust-trapping vortices such as that observed in  the HD~135344B disk can form at the edges of a gap carved by a giant planet \citep{2003ApJ...596L..91K,2007A&A...471.1043D} or at the edges of dead zones \citep[e.g.][]{2006ApJ...649..415I,2012ApJ...756...62L,2015A&A...574A..68F}. In the case of the HD~135344B disk, however, the vortex is not located at the edge of a cavity, but beyond a ring, and may be connected to the spiral arms seen in scattered light. Looking for a consistent explanation for all of the observed substructures can give important information about the number and location of planets potentially embedded in the disk. In particular, are planets needed to explain the observed structures, and which role do the ring and the vortex play in the formation of new planetesimals and planets? On the one hand, the distribution of gas and dust at radii between 0 and 40 au shows that the inner dust cavity is probably due to a planet \citep{vdm2016_isot}, but on the other hand the substructures beyond 40 au are more complex and require a more detailed discussion.

\subsection{Spirals launched by a planet}
The most common explanation for symmetrical spiral arms is an external, massive planet triggering density waves. \citet{2015ApJ...815L..21F,2017ApJ...835...38D} calculate that a $5-10\,\rm M_{\rm J}$ at $r=0.7\asec$ would indeed be able  to excite spiral pressure waves with the shape and the contrast similar to the scattered-light spirals observed in the HD~135344B disk. VLT/SPHERE observations, however, rule out planets with mass $>4\,\rm M_{\rm Jupiter}$  at $r\geq0.6\asec$ \citep{2017A&A...601A.134M}, making this scenario unlikely. These upper limits, though, become higher if cold start formation models are used instead of the hot start ones adopted by \citet{2017A&A...601A.134M}. Moreover, the radial location of the planet is not well constrained by \citet{2015ApJ...815L..21F,2017ApJ...835...38D} and a planet with similar mass, but located at $r\approx0.5\asec$ is still consistent with VLT/SPHERE upper limits even in the hot start scenario. A $5\,\rm M_{\rm J}$ planet located at $r\approx0.5\asec$, i.e., in between the ring and the crescent, could at the same time launch the spiral arms, trigger the vortex asymmetry and carve the gap between the ring and the vortex, explaining all the features at once. In this context, \citet{hammer} also show that a slowly accreting protoplanet can indeed trigger azimuthally asymmetric vortices, such as that observed here which appears to have two different azimuthal widths ($\sigma_{\theta1, \rm V}$ and $\sigma_{\theta2, \rm V}$ in our model).

However, this scenario presents some weaknesses. First, as shown in the right panel of Fig. \ref{fig:data_b4}, the western spiral arm extends further out than the orbit of the planet in this scenario, i.e. $r=0.5\asec$. Assuming an external perturber, the launching point of the spiral would therefore not be consistent with the location of the planet. Second, since both the planet and the vortex rotate around the central star at their local Keplerian velocity, their azimuthal locations are not necessarily the same. The fact that the tip of the western spiral is so close to the vortex center seems therefore to favour a connection between the spirals and the vortex (see Sec. \ref{subsec:vortex}).

In addition, by using the relation between gap width and radial location described in \citet{2016MNRAS.459.2790R,facchini}, it is possible to calculate the mass of the planet carving the observed gap. Assuming a gap with width of $\sim20\,\rm au$ (i.e., $(r_{\rm V}-r_{\rm R})/2$, Tab \ref{tab:bestfit}), located at $\sim68\,\rm au$ from the central star (i.e., $(r_{\rm V}+r_{\rm R})/2$, Tab \ref{tab:bestfit}), and a standard viscosity parameter $\alpha = 10^{-3}$ \citep{1973A&A....24..337S}, the derived mass for the gap-carving planet is only $\sim0.2\,\rm M_{\rm Jupiter}$. This inferred mass is much lower than that required to launch the spiral arms, even though it should be noted that the relation by \citet{2016MNRAS.459.2790R,facchini} was derived for axisymmetric gaps, rather than gaps between a ring and a vortex as in our case.

\subsection{Spirals launched by the vortex}\label{subsec:vortex}
A second possible explanation is that the vortex is massive enough to excite its own spiral density waves. This scenario is theoretically motivated\citep[e.g.][]{2010ApJ...725..146P,2016MNRAS.458.3918Z}, and Fig. \ref{fig:data_b4} indeed shows that the tip of the western, brightest spiral arm is consistent with a launching point close to the Band 4 vortex center. As opposed to planets that can be regarded as point-like, however, vortices are extended sources. No quantitative comparison between models where spirals are launched by vortices or planets has yet been made, and we will here assume that the mass of the spiral-driving-vortex  is comparable to that calculated in the case of an external planet. Using $M_{\rm dust, vortex}\approx50M_\oplus$, as calculated in Section \ref{sec:mass}, and assuming the gas-to-dust ratio of 1 calculated in section \ref{sec:peakshift}, gives a total vortex mass of at least $0.3\,M_{\rm Jupiter}$, which is much lower than that required to launch two symmetrical spiral arms as those observed. 

Given the high uncertainties involved in the calculations of the gas surface density, it is possible to speculate that the gas-to-dust ratio is different than that estimated, and even higher by a factor of 10. Assuming a value of 10 \citep[typical for dust-trapping vortices, e.g.][]{2017ApJ...835..118M, 2017ApJ...850..115S}, the total mass for the vortex is $M_{\rm vortex}\approx1.7\,M_{\rm Jupiter}$. Although still a factor of 3 lower than the mass calculated by \citet{2015ApJ...815L..21F} and \citet{2017ApJ...835...38D}, one should note that the dust mass calculated in section \ref{sec:mass} is only a lower limit (the dust emission in the vortex is only marginally optically thin, even at these long wavelengths), and that the uncertainties involved in the calculations of the dust mass (especially in the dust opacity $\kappa_\nu$) can easily justify the additional missing factor of 3. Moreover, our results also show that dust is trapped and is possibly growing inside the vortex: if planetesimal formation is occurring, a large fraction of the dust mass of the vortex could already be hidden into larger bodies, not contributing to the observed millimeter emission.

If the vortex is not triggered by a planet, the dust morphology observed in the mm-continuum has to be explained by different scenarios. The ring could be what remains of a first generation vortex at the edge of the inner cavity, now decayed into a ring \citep[e.g.][]{2017ApJ...846L...3F,2018MNRAS.tmp.1263P}, that triggered Rossby wave instability developing in a second generation vortex at larger radii \citep{2015ApJ...810...94L}, which is seen in the current observations. Only one planet carving the inner cavity and triggering the first generation vortex would then be required.  This scenario is similar to was proposed by \citet{vdm2016}, with the difference that in the interpretation proposed here both spiral arms, and not only the western one, are due to the vortex.
 
Finally, \citet{2017ApJ...835..118M} have shown that the presence of a ring alongside a vortex is typical when the vortex is generated at the edge of a dead-zone. They also show that dead-zone-generated vortices can overcome the effect of dust feedback, which would tend to destroy the vortex when the gas-to-dust ratio is low. However, in their results the ring is located outside of the vortex (see their Fig. 6 and 7), in contrast with our observations.


\section{Summary and conclusions}
We present new observations of the HD~135344B  transition disk at 155 GHz (ALMA Band 4, 1.9 mm) and  108 GHz (ALMA Band 3, 2.8 mm) and with beam sizes of $0.09\asec\times0.063\asec$ ($12\times9$ au) and  $0.1\asec\times0.073\asec$ ($14\times10$ au), respectively. The high signal-to-noise millimeter continuum maps show a very symmetric inner ring and an asymmetric outer crescent-shaped structure, indicative of a vortex, separated by a narrow gap. This is consistent with the model proposed by \citet{vdm2016} to explain 343 GHz (ALMA Band 7) observations of the same object. The higher signal-to-noise achieved in the Band 4 data also allows us to detect emission from a point source located at the center of the ring, either due to a small circumstellar disk or to free-free emission from hot ionized gas close to the central star.

The Band 3 and Band 4 data have been fitted in the \textit{uv}-plane with a model consisting of an inner Gaussian ring in the radial direction, and an outer double Gaussian both in the radial and azimuthal direction. The same analysis was repeated for archival Band 7 and Band 9 data of the same system. This allows us to quantitatively study the variations in morphology at different wavelengths, and to interpret them in terms of the hydrodynamic response of different grain sizes. Our main conclusions are:
\begin{enumerate}
\item All the four data sets at different wavelengths are well represented by the same model, consisting of a symmetric ring+asymmetric, co-radial crescent.
\item A clear azimuthal variation of the spectal index $\alpha$ (Fig. \ref{fig:alpha}) is found, with the lowest values approaching $\alpha=2$ inside the crescent. This could be due to the fact that the crescent is only marginally optically thin, but is consistent with the presence of large, cm-sized grains \citep{2014prpl.conf..339T}. Future optically thin observations will be needed to disentangle the two effects.
\item A monotonic azimuthal shift in the peak of the emission at different wavelengths is observed for the first time (Fig. \ref{fig:peak_shift}), which cannot be due to Keplerian rotation (opposite direction). Assuming that the emission at longer wavelengths is dominated by larger grains, this shift could mean that the crescent is due to a vortex, and that grains of different size are trapped at different distances from the center of the vortex. A shift of the same magnitude between large and small grains in vortices was indeed predicted by \citet{2016MNRAS.458.3927B}, although in the opposite direction. This indicates that some additional dynamical effect still needs to be studied or that relevant regions of parameter space have yet to be explored.
 \item The emission at longer wavelengths, tracing larger grains, is more azimuthally concentrated than the emission at shorter wavelengths, tracing smaller grains that are more coupled to the gas. This is a key prediction of dust trapping models and simulations \citep{birnstiel2013}.
\item The results summarized above are strong evidence for identifying the crescent-shaped structure of HD~135344B as a vortex. Massive enough vortices can act like planets and trigger spiral density waves. We propose that the vortex is the origin of the spiral density waves observed in scattered light. 
\item A planet of $5\,\rm M_{\rm Jupiter}$ located outside the crescent can produce spiral arms with the shape and contrast of those seen in the HD~135344B  disk \citep{2015ApJ...815L..21F,2015ApJ...809L...5D}. The mass of the vortex, inferred from the dust emission and assuming a gas-to-dust ratio of 10 is $M_{\rm Vortex}=1.7\,\rm M_{\rm Jupiter}$. This value could however be even higher, since the dust emission is only marginally optically thin. A low gas-to-dust ratio 10 or less is consistent with our estimates and with hydrodynamical simulations of gas and dust in vortices \citep{2017ApJ...850..115S,2017ApJ...835..118M}.
\item Our calculations estimate that about half of the dust mass of HD~135344B is trapped in the vortex. Hence,  the vortex is a very favourable spot for planetesimal formation. Planetesimals may already have formed inside the vortex, and could be contributing to its dust mass without showing any emission at mm wavelengths.
\end{enumerate}

Future high resolution observations at shorter wavelengths and optically thin observations at longer ones, together with hydrodynamical modelling to study the proposed scenarios will be crucial to strengthen our interpretation, to confirm the nature of the HD~135344B system and to understand the connection between vortices in the mm and spiral arms in scattered light. Gas observations at higher spatial and spectral resolution will also be critical to constrain the gas content and its kinematics inside the vortex.

\begin{acknowledgements}
We thank  the referee R. Dong for the very constructive comments which improved significantly the clarity of the paper. We also thank 
D. Harsono, Z. Zhu, H. Li, S. Andrews and T. Stolker for very helpful discussions, T. Stolker for making available the SPHERE image of HD~135344B, and the ALMA director for granting DDT time. P.P. acknowledges support by NASA through Hubble Fellowship grant HST-HF2-51380.001-A awarded by the Space Telescope Science Institute, which is operated by the Association of Universities for Research in Astronomy, Inc., for NASA, under contract NAS 5-26555. M.B. acknowledges funding from ANR of France under contract number ANR-16-CE31-0013 (Planet Forming disks). M.T. has been supported by the DISCSIM project, grant agreement 341137 funded by the European Research Council under ERC-2013-ADG. Astrochemistry in Leiden is supported by the European Union A-ERC grant 291141 CHEMPLAN,
by the Netherlands Research School for Astronomy (NOVA), and by a Royal
Netherlands Academy of Arts and Sciences (KNAW) professor prize. This paper makes use of the following ALMA data: 2011.0.00724.S, 2012.1.00158.S, 2016.1.00340.S, 2017.1.00884.S and 2017.A.00025.S. ALMA is a partnership of ESO (representing its member states), NSF (USA) and NINS (Japan), together with NRC (Canada) and NSC and ASIAA (Taiwan) and KASI (Republic of Korea), in cooperation with the Republic of Chile. The Joint ALMA Observatory is operated by ESO, AUI/NRAO and NAOJ.All the figures
were generated with the \texttt{python}-based package \texttt{matplotlib} \citep{hunter2007}.
\end{acknowledgements}


\begin{appendix}
\section{Fit results}\label{sec:appendix_visibilities}
Here we report the results of the fits for the four different data sets. For each data set we show a triangle plot, showing the MCMC results and the marginalized posteriors for each parameter and the deprojected real and imaginary part of the models (red lines) obtained with the parameters in Tab. \ref{tab:bestfit} compared to the data (black dots). For Band 3 and Band 4, only the inner $1000\,\rm k\lambda$ are shown.
\begin{figure*}
\centering
\includegraphics[width=\textwidth]{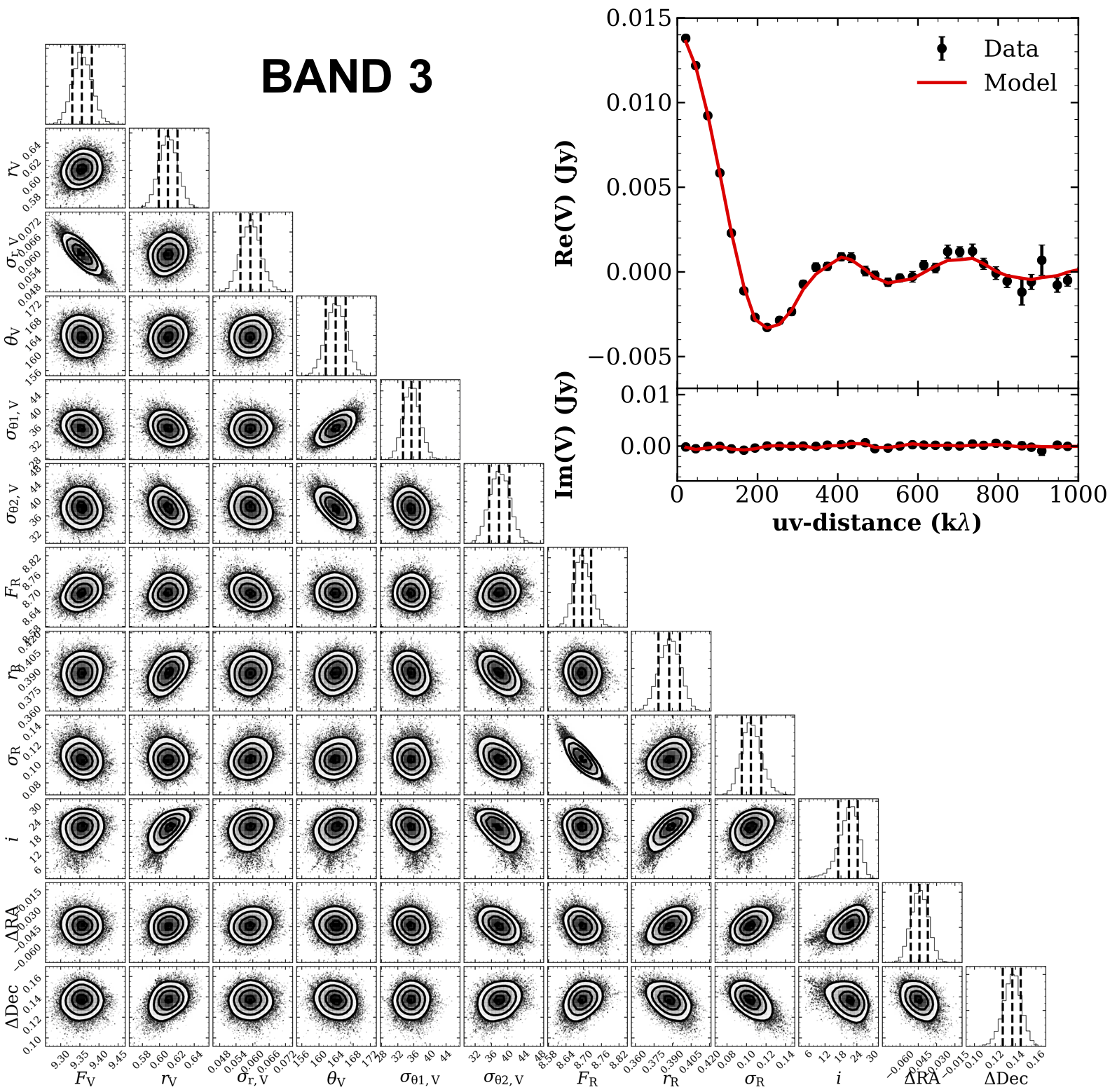}
\caption{Fit results for the Band 3 data set.}
\end{figure*}
\begin{figure*}
\centering
\includegraphics[width=\textwidth]{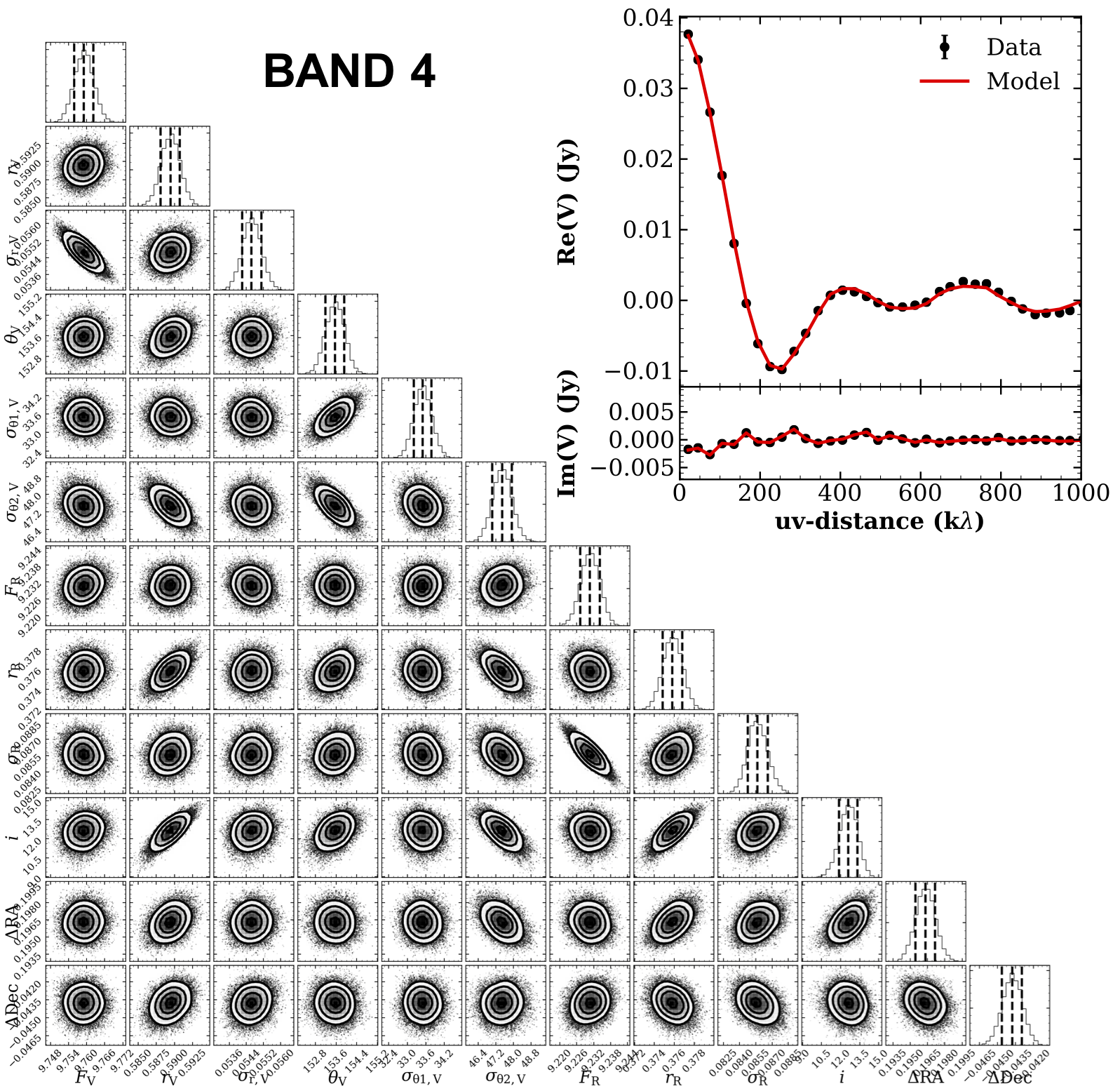}
\caption{Fit results for the Band 4 data set.}
\end{figure*}
\begin{figure*}
\centering
\includegraphics[width=\textwidth]{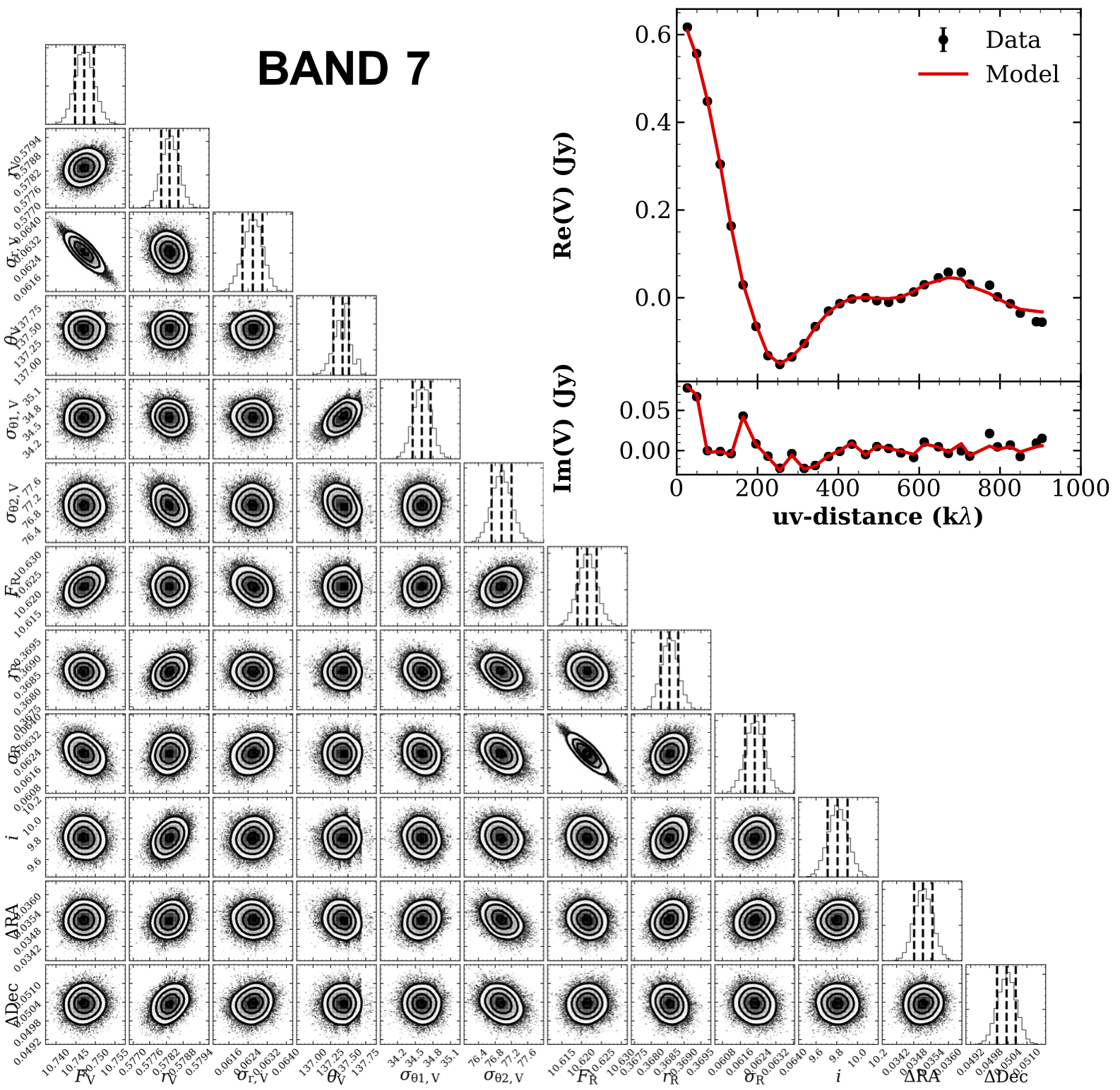}
\caption{Fit results for the Band 7 data set.}
\end{figure*}
\begin{figure*}
\centering
\includegraphics[width=\textwidth]{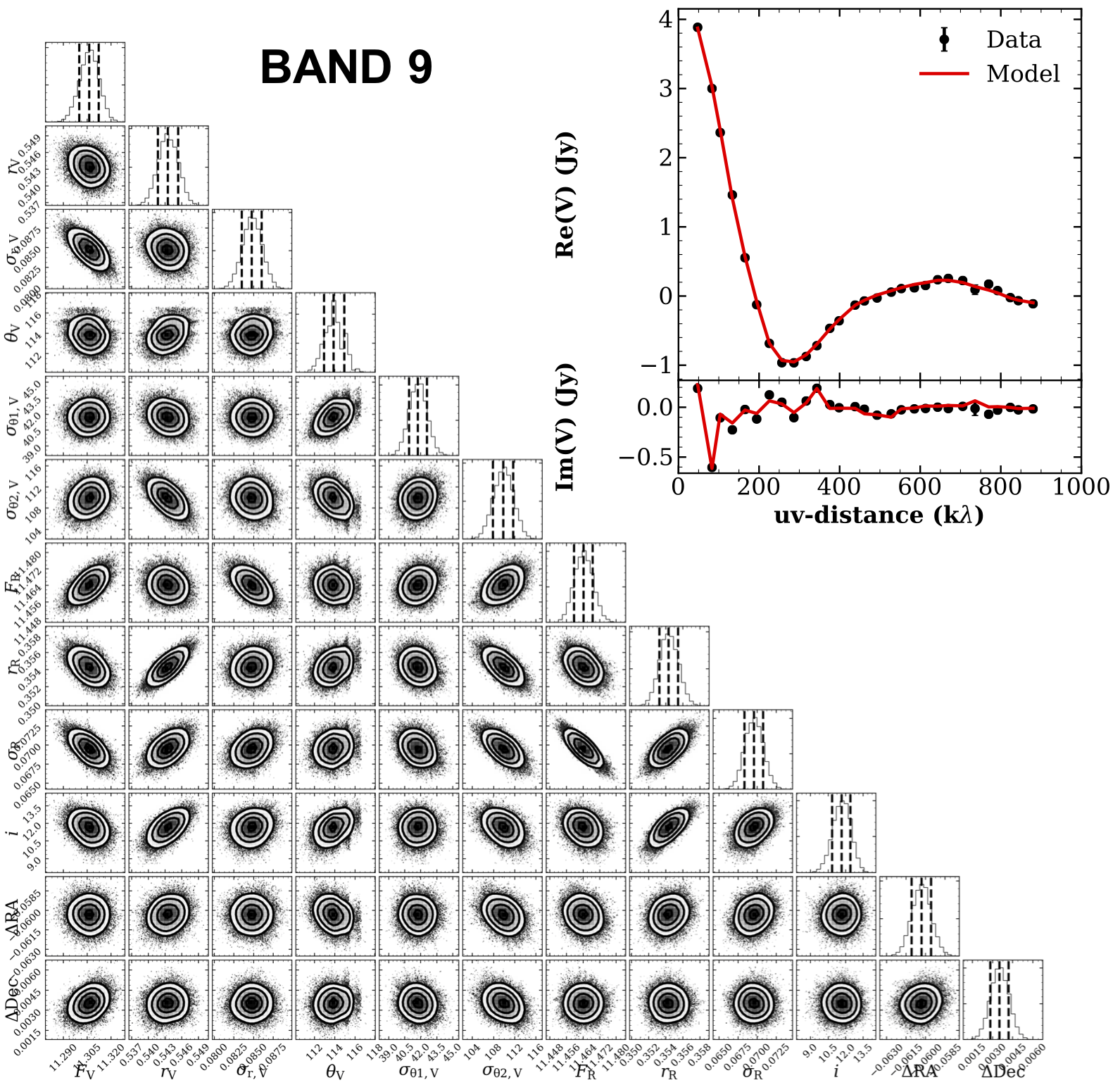}
\caption{Fit results for the Band 9 data set.}
\end{figure*}

\end{appendix}

\end{document}